\def\rfr#1{eq. (\ref{#1})}

\def\dert#1#2{\frac{{{d}}{#1}}{{{d}}{#2}}}

\def\eqi{\begin{equation}}
\def\eqf{\end{equation}}
\def\eqia{\begin{eqnarray}}
\def\eqfa{\end{eqnarray}}
\def\Om{\mathit{\Omega}}
\def\rp#1#2{{#1\over#2}} \def\lb#1{\label{#1}}
\def\kap{\bds{\hat{k}}}
\def\kx{\hat{k}_x}
\def\ky{\hat{k}_y}
\def\kz{\hat{k}_z}
\def\bds#1{\boldsymbol{#1}}

\def\coo{\cos 2\omega}
\def\soo{\sin 2\omega}
\def\cO{\cos\Om}
\def\sO{\sin\Om}
\def\cOO{\cos 2\Om}
\def\sOO{\sin 2\Om}

\def\cI{\cos I}
\def\sI{\sin I}
\def\cII{\cos 2I}
\def\sII{\sin 2I}

\def\ton#1{\left(#1\right)}
\def\qua#1{\left[#1\right]}
\def\grf#1{\left\{#1\right\}}
\def\ang#1{\left\langle #1\right\rangle}

\documentclass[11pt]{article}
\usepackage{url}
\usepackage{amsmath,amsthm,amscd,amssymb}
\usepackage{latexsym,w-greek,wasysym}
\usepackage{graphicx,epsfig}
\usepackage{hyperref}
\allowdisplaybreaks[1]

\RequirePackage{color}

\begin{document}

\title{\textcolor{black}{Lower bounds of characteristic scale of topological modification of
the Newtonian gravitation}}

\author{L. Iorio \\ Ministero dell'Istruzione, dell'Universit\`{a} e della Ricerca (M.I.U.R.)-Istruzione \\ Fellow of the Royal Astronomical Society (F.R.A.S.) \\
 International Institute for Theoretical Physics and
Advanced Mathematics Einstein-Galilei \\ Permanent address: Viale Unit$\grave{\rm a}$ di Italia 68
70125 Bari (BA), Italy \\ email: lorenzo.iorio@libero.it}

\maketitle

\begin{abstract}
We analytically work out the long-term orbital perturbations induced by the \textcolor{black}{leading order} of  perturbing potential arising from the local modification of the Newton's inverse square law due to a topology $\mathbb{R}^2\times \mathbb{S}^1$ with a compactified dimension of radius $R$ recently proposed by Floratos and Leontaris. We neither restrict to any specific spatial direction $\kap$ for the asymmetry axis nor to particular orbital configurations of the test particle. Thus, our results are quite general. Non-vanishing long-term variations  occur for all the usual osculating Keplerian orbital elements, apart from the semimajor axis which is left unaffected. By using  recent improvements in the determination of the orbital motion of Saturn from Cassini data, we preliminarily inferred  $R\gtrsim 4-6\ {\rm kau}$. As a complementary approach, the putative topological effects should be explicitly modeled and solved-for with a modified version of the ephemerides dynamical models with which the same data sets should be reprocessed.
\end{abstract}

%
\centerline
{PACS: 04.80.-y; 04.80.Cc; 04.50.Kd; 95.10.Km}
\section{Introduction}
In \cite{Flo012} non-trivial topological modifications of the Newton's law were proposed on the ground of certain considerations of astrophysical and cosmological nature pertaining various forms of anisotropies occurring at such scales. In particular, Floratos and Leontaris \cite{Flo012} looked at local changes of the topology of the Euclidean space from $\mathbb{R}^3$ to $\mathbb{R}^2\times \mathbb{S}^1$, with a compactified dimension with scale $R$.

In Section \ref{calcoli}, we  will analytically work out the orbital effects induced by them on the motion of a test particle moving around a central body of mass $M$. For distances $r$ smaller\footnote{\textcolor{black}{Floratos and Leontaris \cite{Flo012} discussed also the case $r\gg R$.}} than the compactification radius $R$, the correction to the Newtonian potential $U_{\rm N}$ is \cite{Flo012}
\eqi
U_{\rm pert} = -\rp{2GM}{r}\sum_{j=1}^{\infty}\zeta\ton{2j+1}\ton{\rp{r}{2\pi R}}^{2j+1}P_{2j}\ton{\cos\theta},\lb{upert}
\eqf
where $G$ is the Newtonian constant of gravitation, $\zeta$ is the Riemann Zeta function,  $P_{2j}$ is the Legendre polynomial of degree $2j$, and $\cos\theta$ is the cosine of the angle between the unit position vector $\bds{\hat{r}}$ of the point in space at which the potential is evaluated and the direction $\bds{\hat{k}}$ of the asymmetry axis. Contrary to Floratos and Leontaris \cite{Flo012},  we will neither align $\bds{\hat{k}}$ with the reference $z$ axis nor restrict to any specific orbital configuration for the test particle.

Since it is expected that $R$ is quite large, in the following we will consider only the effects induced by the \textcolor{black}{leading order term  $(j=1)$} in \rfr{upert}. Such an assumption will be  a-posteriori justified in Section \ref{osservazioni}, where we preliminarily compare our results to the latest observational determinations of the orbital motion of Saturn.

Section \ref{conclusioni} is devoted to the conclusions.
\section{The long-term rates of change of the osculating Keplerian orbital elements}\lb{calcoli}
The long-term rates of change of the usual osculating Keplerian orbital elements due to \textcolor{black}{the leading order term of $U_{\rm pert}$ ($j=1$)} can be straightforwardly worked out with standard perturbative techniques. For example, by averaging the \textcolor{black}{leading order} term of \rfr{upert} over one orbital revolution, the Lagrange equations for the variation of the elements \cite{roy05} yield
{\begin{align}
\ang{\dert a t} \lb{dadt} & = 0, \\ \nonumber \\
\ang{\dert e t} \lb{dedt} & = -\rp{15\zeta\ton{3}a^3 n_{\rm b}e\sqrt{1 - e^2}}{16\pi^3 R^3}{\mathfrak{E}}, \\ \nonumber \\
\mathfrak{E} \nonumber & = \cos^2 I \sin 2\omega \ton{\ky \cos \Om - \kx \sin \Om}^2 + 2 \kz \cos 2\omega \sin I \ton{\kx \cos \Om + \ky \sin \Om} -\\ \nonumber \\
\nonumber & -  \rp{1}{2} \sin 2\omega \qua{\kx^2 + \ky^2 - 2\kz^2\sin^2 I + \ton{\kx^2 - \ky^2}\cos 2\Om + 2 \kx \ky \sin 2\Om} + \\ \nonumber \\
\nonumber & + \cos I \grf{2 \kz \sin I \sin 2\omega \ton{\ky \cos \Om - \kx \sin \Om} + \right.\\ \nonumber \\
& + \left. \cos 2\omega \qua{2 \kx \ky \cos 2\Om + \ton{\ky^2 -\kx^2} \sin 2\Om}}, \\ \nonumber \\
\ang{\dert I t} \lb{dIdt} & = \rp{3\zeta\ton{3}a^3 n_{\rm b}}{16\sqrt{1-e^2}\pi^3 R^3}{\mathfrak{I}}, \\ \nonumber \\
\mathfrak{I} \nonumber & = \qua{\kz \cI + \sI \ton{\kx \sO -\ky \cO}} \grf{5 e^2 \soo\qua{\kz \sI  + \right.\right. \\ \nonumber \\
\nonumber & + \left.\left.  \cI  \ton{\ky \cO - \kx \sO}} + \right. \\ \nonumber \\
& + \left. 2\qua{1 + \rp{3}{2} e^2\ton{1 + \rp{5}{3}\coo}} \ton{\kx \cO + \ky \sO}}, \\ \nonumber \\
\ang{\dert \Om t} \lb{dOdt} & = -\rp{3\zeta\ton{3}a^3 n_{\rm b}\csc I}{16\sqrt{1-e^2}\pi^3 R^3}{\mathfrak{O}}, \\ \nonumber \\
\mathfrak{O} \nonumber & = \qua{\kz \cI + \sI \ton{\kx \sO -\ky \cO}} \grf{-2 \kz \sI - \right. \\ \nonumber \\
\nonumber & - \left. 2 \cI \qua{1 + \rp{3}{2} e^2\ton{1 - \rp{5}{3} \coo}} \ton{\ky \cO - \kx \sO} + \right. \\ \nonumber \\
& + \left. e^2 \qua{\kz \ton{5 \coo - 3} \sI - 5 \soo \ton{\kx \cO + \ky \sO}}}, \\ \nonumber \\
\ang{\dert \varpi t} \lb{dodt} & = \rp{3\zeta\ton{3} a^3 n_{\rm b}}{32\sqrt{1 - e^2}\pi^3 R^3}\mathfrak{P}, \\ \nonumber \\
\mathfrak{P} \nonumber & =  \ton{1 - e^2} \grf{-4 + 2 \sin^2\omega \qua{\kx^2 + \ky^2 + 4 \kz^2 + 2 \ton{\kx^2 + \ky^2 - 2 \kz^2} \cII - \right.\right. \\ \nonumber \\
\nonumber & - \left.\left. \ton{\kx^2 - \ky^2} \ton{3 + 2 \cII} \cOO + 8 \ky \kz \cO \sII - 8 \kx \kz \sII \sO} + \right.\\ \nonumber \\
\nonumber & + \left. \cos^2\omega \qua{7 \ton{\kx^2 + \ky^2} - 2 \kz^2 - \ton{\kx^2 + \ky^2 - 2 \kz^2} \cII + \right.\right.\\ \nonumber \\
\nonumber & + \left.\left.\ton{\kx^2 - \ky^2} \ton{9 + \cII} \cOO - 4 \ky \kz \cO \sII + 4 \kx \kz \sII \sO} + \right. \\ \nonumber \\
\nonumber & + \left. 20 \soo \qua{\kx \ky \cI \cOO + \kz \sI \ton{\kx \cO + \ky \sO}} - \right.\\ \nonumber \\
\nonumber & - \left. \qua{-5 \kx \ky \ton{3 + \cII} \coo - 6 \kx \ky \sin^2 I + 10 \ton{\kx^2 - \ky^2}  \cI \soo} \sOO} - \\ \nonumber \\
\nonumber & - 2 \qua{\kz \cI + \sI \ton{\kx \sO -\ky \cO}} \grf{-2 \kz \sI - \right.\\ \nonumber \\
\nonumber & -\left. 2 \cI \qua{1 + \rp{3}{2}e^2\ton{1 - \rp{5}{3} \coo}} \ton{\ky \cO - \kx \sO} + \right. \\ \nonumber \\
& + \left. e^2 \qua{\kz \ton{5 \coo - 3} \sI - 5 \soo \ton{\kx \cO + \ky \sO}}} \tan\ton{\rp{I}{2}}, \\ \nonumber \\
\ang{\dert {\mathcal{M}} t} - n_{\rm b} \lb{dMdt} & = \rp{\zeta\ton{3} a^3 n_{\rm b}}{64\pi^3 R^3}\mathfrak{M}, \\ \nonumber \\
\mathfrak{M} \nonumber & = -13 \ton{1 + \rp{12}{13} e^2} \ton{-8 + 9 \kx^2 + 9 \ky^2 + 6 \kz^2} -\\ \nonumber \\
\nonumber & -  120 \ton{1 + 4 e^2} \soo \ton{\kx \cO + \ky \sO} \qua{\kz \sI + \cI \ton{\ky \cO - \kx \sO}} + \\ \nonumber \\
\nonumber & +  15 \ton{1 + 4 e^2} \coo \qua{3 \ton{\kx^2 - \ky^2} \cOO + 2 \ton{\kx^2 + \ky^2 - 2 \kz^2} \sin^2 I -\right.\\ \nonumber \\
\nonumber & - \left. 4 \kz \sII \ton{\ky \cO - \kx \sO} + 6 \kx \ky \sOO} +\\ \nonumber \\
\nonumber & +  78 \ton{1 + \rp{12}{13} e^2} \qua{\ton{\kx^2 - \ky^2} \cOO \sin^2 I + 2 \kz \sII \ton{\ky \cO - \kx \sO} +\right.\\ \nonumber \\
\nonumber & + \left. 2 \kx \ky \sin^2 I \sOO} + 3\cII \grf{13\ton{1 + \rp{12}{13} e^2} \ton{\kx^2 + \ky^2 - 2 \kz^2} + \right.\\ \nonumber \\
& +\left. 5 \ton{1 + 4 e^2} \coo \qua{\ton{\kx^2 - \ky^2} \cOO + 2 \kx \ky \sOO}},
\end{align}
}
where \cite{roy05} $a,e,I,\Om,\omega,\varpi\doteq\Om + \omega,\mathcal{M}$ are the semimajor axis, the eccentricity, the inclination to the reference $\{x,y\}$ plane of the coordinate system adopted, the longitude of the ascending node, the argument of pericenter, the longitude of pericenter and the mean anomaly, respectively;  $n_{\rm b}\doteq \sqrt{GM/a^3}$ is the Keplerian mean motion.  \textcolor{black}{The simplifying condition $e,I\sim 0$  was not imposed} in obtaining \rfr{dadt}-\rfr{dMdt}, which are, thus, exact.

According to \rfr{dedt}, for circular orbits ($e=0$) the eccentricity is left unaffected.
\section{Confrontation with the observations}\lb{osservazioni}
By using \rfr{dadt}-\rfr{dMdt} it is possible to infer lower bounds on $R$ from \textcolor{black}{solar system} planetary orbital motions. Recently, Fienga et al. \cite{fienga011} determined supplementary precessions of the perihelia and the nodes of some planets of our solar system which, in principle, account for any unmodeled/mismodeled dynamical effects with respect to those accounted for in the force models fitted to the observations in producing the INPOP10a ephemerides which use the mean terrestrial equator at J$2000.0$ as reference $\{x,y\}$ plane. Since the predicted rates of change are proportional to $a^{3/2}$, the largest effects are sensed by the outer planets. Thus, we will use the perihelion and the node of Saturn, whose precessions are nowadays known at a $\sim 0.4-0.6$ mas cty$^{-1}$ level of accuracy \cite{fienga011}.

In Figure \ref{rcomp} we display the lower bounds on $R$, in kau, obtained from the Kronian perihelion and node as functions of the right ascension $\alpha$ and declination $\delta$ of the asymmetry axis $\kap$.
\begin{figure}
\centering
\begin{tabular}{cc}
\epsfig{file=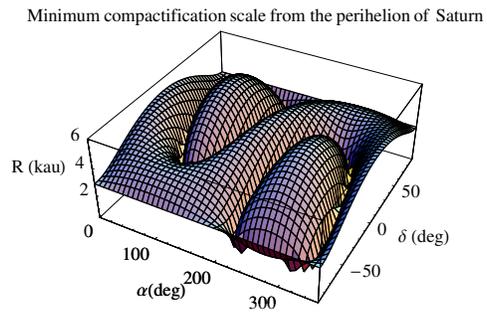,width=0.40\linewidth,clip=} \\
\epsfig{file=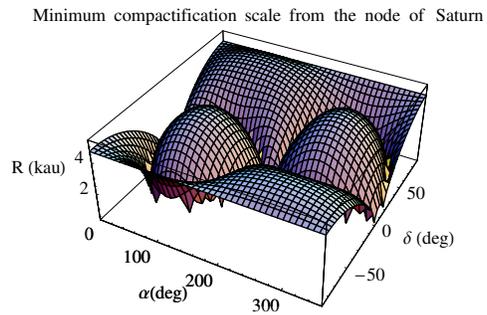,width=0.40\linewidth,clip=}
\end{tabular}
\caption{Minimum compactification scale $R$ from the supplementary advances of the perihelion and the node of Saturn by Fienga et al. \cite{fienga011} as a function of the right ascension $\alpha$ and the declination $\delta$ of the asymmetry axis $\kap$. We adopted \rfr{dodt} and \rfr{dOdt}.}\lb{rcomp}
\end{figure}
\textcolor{black}{From an inspection of Figure \ref{rcomp}, it turns out that  the perihelion of Saturn yields up to $R\gtrsim 6$ kau for most of locations in the sky of the asymmetry axis.  The perihelion-inferred bounds on $R$ are weaker in a few positions of $\bds{\hat{k}}$, but the node can  be fruitfully  used in a complementary way to infer tight constraints also there. Thus, we can reasonably conclude that $R\gtrsim 4-6$ kau for all over the sky.}
\section{Conclusions}\lb{conclusioni}
We looked at some potentially observable consequences of a possible non-trivial local modification $\mathbb{R}^2 \times \mathbb{S}^1$ of the topology of the Euclidean space with a compactified spatial dimension with scale $R$ recently proposed by Floratos and Leontaris. We analytically worked out the orbital effects due to the first term in the expansion of the resulting perturbing potential. Apart from the semimajor axis, all the other standard osculating Keplerian orbital elements undergo non-vanishing long-term variations. We did not make  a-priori simplifying assumptions on the orbital configuration of the perturbed test particle. Moreover, we did not choose any specific spatial direction for the anisotropy axis.

We preliminarily inferred lower bounds on $R$ from the latest improvements in the determination of the orbital motion of Saturn from Cassini data. We obtained $R\gtrsim 4-6\ {\rm kau}$, \textcolor{black}{depending on the orientation of the asymmetry axis in the sky}. With such figures it is possible, a posteriori, to justify our
initial choice of considering just the \textcolor{black}{leading order} term in the expansion of the perturbing potential since the precessions of higher order, computed for them, would be far smaller that the present-day level of accuracy in determining the extra-rates of the node and the perihelia of Saturn. We did not consider other known mismodeled/unmodeled forces causing potentially competing precessions like the Sun's oblateness, \textcolor{black}{the PPN $\beta$ parameter} and the Lense-Thirring effect since, \textcolor{black}{at present}, they are negligible for Saturn \textcolor{black}{with respect to the current level of accuracy in constraining its supplementary precessions}.

For a more refined analysis, it would be possible, in principle,  to employ an ad-hoc modified version of the usual suite of standard force models to be fitted to the same data records, and explicitly solve for $R$ or some other equivalent adjustable free parameter.

As a possible further investigation of the viability of the suggested modified topology, the behavior of a typical Oort comet may be studied in order to see if the usual picture of the Oort cloud,  which extends well beyond the previously inferred lower bound for $R$, would be notably altered.
\bibliography{Topologicalbib}{}

\providecommand{\href}[2]{#2}\begingroup\raggedright\begin{thebibliography}{1}

\bibitem{Flo012}
E.~Floratos and G.~Leontaris, ``On topological modifications of newton's law,''
  \href{http://arxiv.org/abs/1202.6067}{{\ttfamily arXiv:1202.6067
  [astro-ph.CO]}}.

\bibitem{roy05}
A.~E. Roy, {\em Orbital Motion}.
\newblock Institute of Physics, Bristol, fourth edition~ed., 2005.

\bibitem{fienga011}
A.~Fienga, J.~Laskar, P.~Kuchynka, H.~Manche, G.~Desvignes, M.~Gastineau,
  I.~Cognard, and G.~Theureau, ``The inpop10a planetary ephemeris and its
  applications in fundamental physics,'' {\em Celestial Mechanics and Dynamical
  Astronomy} {\bfseries 111} no.~3, (2011) 363--385.

\end{thebibliography}\endgroup

\end{document}